\documentclass[aps,preprint,superscriptaddress]{revtex4}
\usepackage{graphicx,subfigure,textcomp}
\usepackage{amsthm}
\usepackage{comment}
\usepackage{rotating}
\usepackage{amsmath,amssymb,amsfonts}
\usepackage{wasysym} 
\usepackage{pifont} 
\usepackage{multirow}
\usepackage{tabularx}
\usepackage{setspace}
\usepackage{color,float}
\usepackage{adjustbox,natbib}

\newcounter{reaction} 
\usepackage[%
  colorlinks=true,
  urlcolor=blue,
  linkcolor=blue,
  citecolor=blue
]{hyperref}

\begin{document}
\bibliographystyle{revtex4-1}


\title{The role of sputtered atom and ion energy distribution in films deposited by Physical Vapor Deposition: A molecular dynamics approach}



\author{Soumya Atmane}
\affiliation{GREMI, UMR7344 CNRS Universit{\'e} d’Orl{\'e}ans, Orl{\'e}ans, France}

\author{Alexandre Maroussiak}
\affiliation{GREMI, UMR7344 CNRS Universit{\'e} d’Orl{\'e}ans, Orl{\'e}ans, France}

\author{Ama{\"e}l Caillard}
\affiliation{GREMI, UMR7344 CNRS Universit{\'e} d’Orl{\'e}ans, Orl{\'e}ans, France}

\author{Anne-Lise Thomann}
\affiliation{GREMI, UMR7344 CNRS Universit{\'e} d’Orl{\'e}ans, Orl{\'e}ans, France}

\author{Movaffaq Kateb}
\affiliation{Condensed Matter and Materials Theory, Department of Physics, Chalmers University, Sweden}

\author{Jon Tomas Gudmundsson}
\affiliation{Science Institute, University of Iceland,
Dunhaga 3, IS-107 Reykjavik, Iceland}
\affiliation{Division of Space and Plasma Physics, School of Electrical Engineering and Computer Science, KTH Royal Institute of Technology, SE-10044  Stockholm, Sweden}

\author{Pascal Brault}
\affiliation{GREMI, UMR7344 CNRS Universit{\'e} d’Orl{\'e}ans, Orl{\'e}ans, France}
\affiliation{MS4ALL, Lab’O Village by CA, Orl{\'e}ans, France}
 \email{pascal.brault@univ-orleans.fr; pascal.brault@ms4all.eu}


\date{\today}

\begin{abstract} 
We present a comparative study of copper film growth with a constant energy neutral beam, thermal evaporation, dc magnetron sputtering, high-power impulse magnetron sputtering (HiPIMS), and bipolar HiPIMS, through molecular dynamics simulations. Experimentally determined energy distribution functions were utilized to model the deposition processes. Our results indicate significant differences in the film quality, growth rate, and substrate erosion between the various physical vapor deposition techniques. Bipolar HiPIMS shows the potential for improved film structure under certain conditions, albeit with increased substrate erosion.  Bipolar +180 V HiPIMS with 10\% Cu$^+$ ions exhibited the best film properties in terms of crystallinity and atomic stress among the PVD processes investigated.

\end{abstract}

\maketitle


Molecular dynamics (MD) simulation is a simulation technique well suited for describing thin film or nanoparticle growth phenomena \citep{xie14:224004,neyts17:1600145,brault23:19, brault24:1}.  Here, MD simulations are applied to compare the growth of thin copper (Cu) films using various physical vapor deposition (PVD) techniques \citep{gudmundsson22:083001}.  
The deposition systems assumed in the present work, include a constant energy neutral beam, thermal evaporation, dc magnetron sputtering (dcMS), high power impulse magnetron sputtering (HiPIMS), and bipolar HiPIMS \citep{gudmundsson20:113001,gudmundsson22:083001}. In neutral beam, thermal evaporation, and dcMS depositions the film forming species are neutral atoms.  In HiPIMS operation the magnetron sputtering discharge is driven by high power pulses at low frequency and short duty cycle \citep{gudmundsson20:113001,gudmundsson12:030801} and the ionized flux fraction of the sputtered species can be significant 
\citep{fischer23:125006}, and furthermore the ion energy can be significantly higher than in dcMS operation \citep{bohlmark06:1522}.    In bipolar HiPIMS operation a positive voltage is applied to the target following the negative sputter pulse, which increases the plasma potential and shifts the ion energy distribution (IED) to higher energies \citep{keraudy19:433,michiels:415202,walk:065002,zanaska:025007}. 
 Some successful attempts have been made to include the specific characteristics
of different magnetron sputtering (MS) deposition techniques into molecular dynamics simulations,  the role of the ionization fraction \citep{kateb19:031306}, the ion potential energy \citep{kateb21:127726}, and the effect of substrate bias \citep{kateb20:043006}, which show varying film properties depending on the deposition method.

Here, we further improve the description of the physically released film forming species by using selected initial velocity conditions from experimental data (energy-resolved mass spectrometry) and SRIM simulations \citep{biersack84:73}  for the  ion energy distribution and the  atom energy distribution (AED).
This choice was motivated by approaching the experimental conditions as closely as possible. Usually, in MD simulations, this is achieved by setting a mean kinetic energy or sampling from a uniform energy distribution.
Further improvement of the initial conditions is expected to give a step forward for better MD simulation prediction of sputtered film properties, such as morphology, composition, structure, and tribology.

\begin{figure}[htp!]
\begin{center}
\includegraphics[width=3 in]{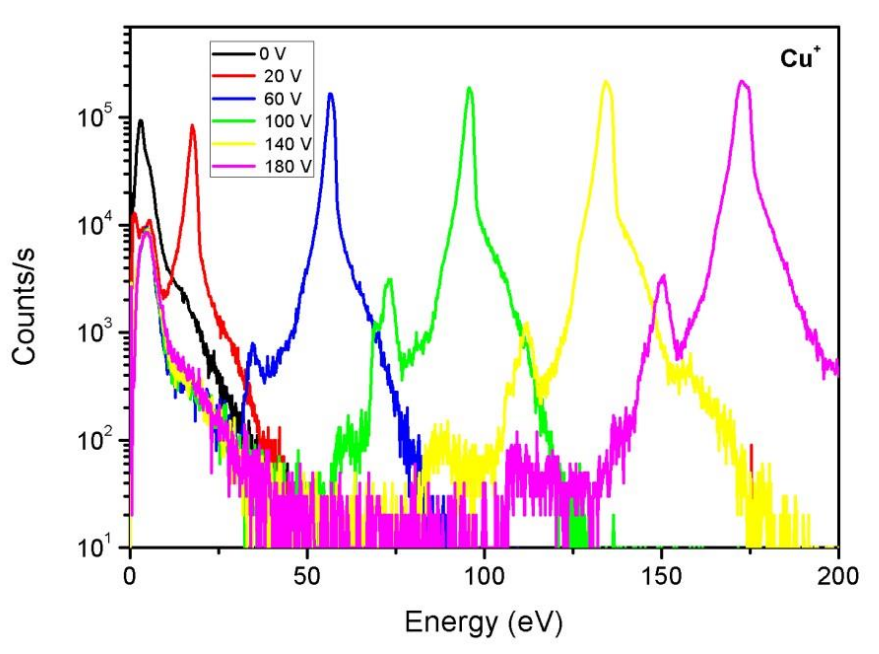}
 \caption{ Experimentally determined copper ion Cu$^+$ ion energy distributions (IEDs) (semilog scale) in HiPIMS mode (0V) and bipolar HiPIMS for various positive voltages recorded 10 cm away from the target surface. }
 \label{fig:IEDBHiPIMS}
\end{center}
\end{figure}

The reactor used to determine the IED for the magnetron sputtering processes 
is made up of a central chamber, which is a cylindrical stainless steel enclosure with a diameter of 200 mm. It contained six crossed flanges with an external diameter of 240 mm (ISO-K200). The first flange was connected to a pumping system that ensured a vacuum of approximately $5 \times 10^{-5}$ Pa using a primary pump (Pfeiffer vacuum, ACP15, 15 m$^3$/h) and a turbo-molecular pump (Pfeiffer vacuum, 500 l/s). The lamination valve placed between the chamber and the turbo-molecular pump made it possible to adjust the working gas pressure inside the enclosure, and the two gauges (a combined pirani/cold cathode gauge and baratron gauge) made it possible to measure this pressure. The magnetron assembly (Angstrom Sciences (ONYX-2)) where the sputtering target (copper) is placed, had a diameter of 2 inches. 
It was placed facing the EQP1000 300 amu Hiden Analytical mass spectrometer at a distance of 10 cm. A pulsed power module (Starfire Industries), that contains a solid-state switching device, was fed by a Kurt J. Lesker PDX 500 DC generator.  
This system can be used to power the magnetron sputtering discharge with unipolar pulses (HiPIMS) or bipolar pulses (bipolar-HiPIMS) with a voltage limit of 1000 V. The current-voltage characteristics were observed using a Tektronix oscilloscope. Argon was used as the working gas, set at 0.7 Pa, and the flow rate was regulated at 40 sccm using a Bronkhorst mass flow meter (F-200CV).   In the present work, the HiPIMS sputter pulse was 50 $\mu$s long and the negative bias  -650 V. In bipolar HiPIMS configuration, the negative pulse was followed by a longer pulse (250 $\mu$s) with a positive bias varying from +20 to +180 V.  Figure \ref{fig:IEDBHiPIMS} shows the corresponding experimentally recorded IEDs for Cu$^+$ ions measured 10 cm from the Cu target center.

\begin{figure}[htp!]
\begin{center}
\includegraphics[width=3 in]{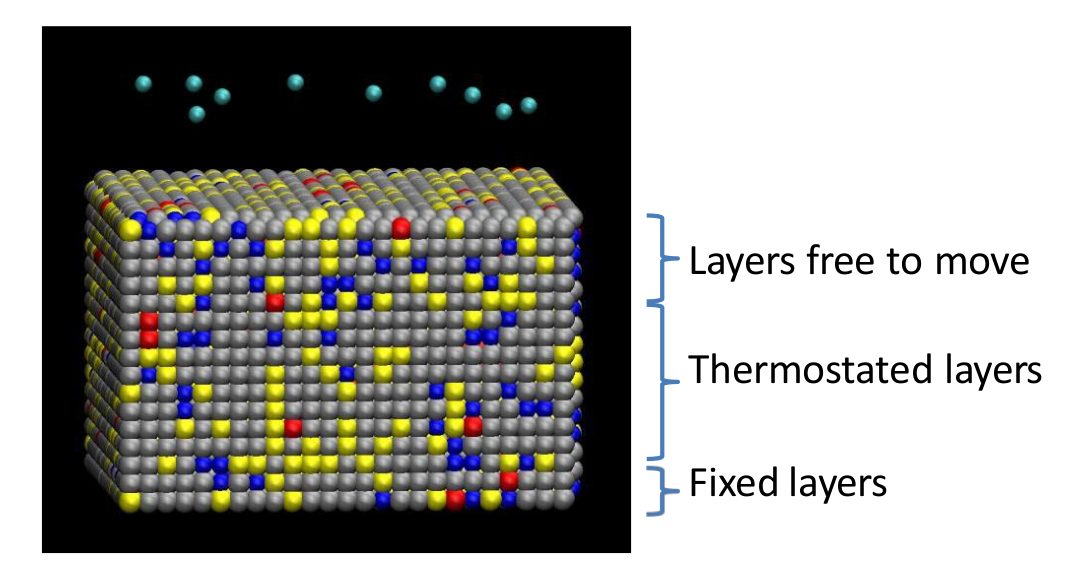}
 \caption{ Simulation box with the stainless steel substrate and the first set of 10 atoms/ions before release
towards the surface. Color code: cyan Cu, yellow Cr, silver Fe, red Mo, blue Ni.}
 \label{fig:simulationbox}
\end{center}
\end{figure}

Molecular dynamics simulations allow the calculation of the trajectories of a set of species by solving Newton’s equations of motion:
\begin{equation}
    m_i \frac{{\rm d}^2\vec{r}_i(t)}{{\rm d}t^2} = \vec{F}_i = - \vec{\nabla} V(\vec{r}_1(t),\vec{r}_2(t),\dots,\vec{r}_N(t))
\end{equation} 
where $\vec{r}_i(t)$ is the position of atom $i$ at time $t$ with mass $m_i$, and $V$ is the interaction potential between all
$N$ involved species. These equations only require knowledge of two initial conditions, positions and velocities for all species at the initial time $t = 0$, and of the interactions between all species at all  times. The initial
positions refer to the geometry/topology of the particles at the beginning of the simulation, while velocities
are (randomly) selected from a velocity distribution that is consistent with the deposition method under study. To examine the effect of the AEDs and IEDs on the deposited films, we considered a stainless-steel substrate
(bcc Fe$_{67}$Cr$_{17}$Mo$_{2}$Ni$_{14}$, $72 \times 72 \times 46$ \AA$^3$ slab composed of 20000 atoms).  For a constant energy neutral beam, thermal evaporation, and dcMS 10000 neutral Cu atoms are periodically released towards the surface.  In the case of HiPIMS and bipolar HiPIMS, 10000 species composed of 90\% Cu + 10\% Cu$^+$ or 50\% Cu + 50\% Cu$^+$ were released at a
periodic rate from random positions above the surface. Such neutral and ion compositions are chosen for mimicking low and high ion flux towards the substrate. Sets of 10 Cu (or $x$Cu + $y$Cu$^+$) atoms, randomly located above the surface and sufficiently far from each other, were released every 40 ps. Thus, the total
simulation time was 40 ns. Figure \ref{fig:simulationbox} shows the initial simulation box.

The initial velocities are randomly selected from the corresponding AEDs and IEDs.
\begin{figure}[htp!]
\begin{center}
\includegraphics[width=5 in]{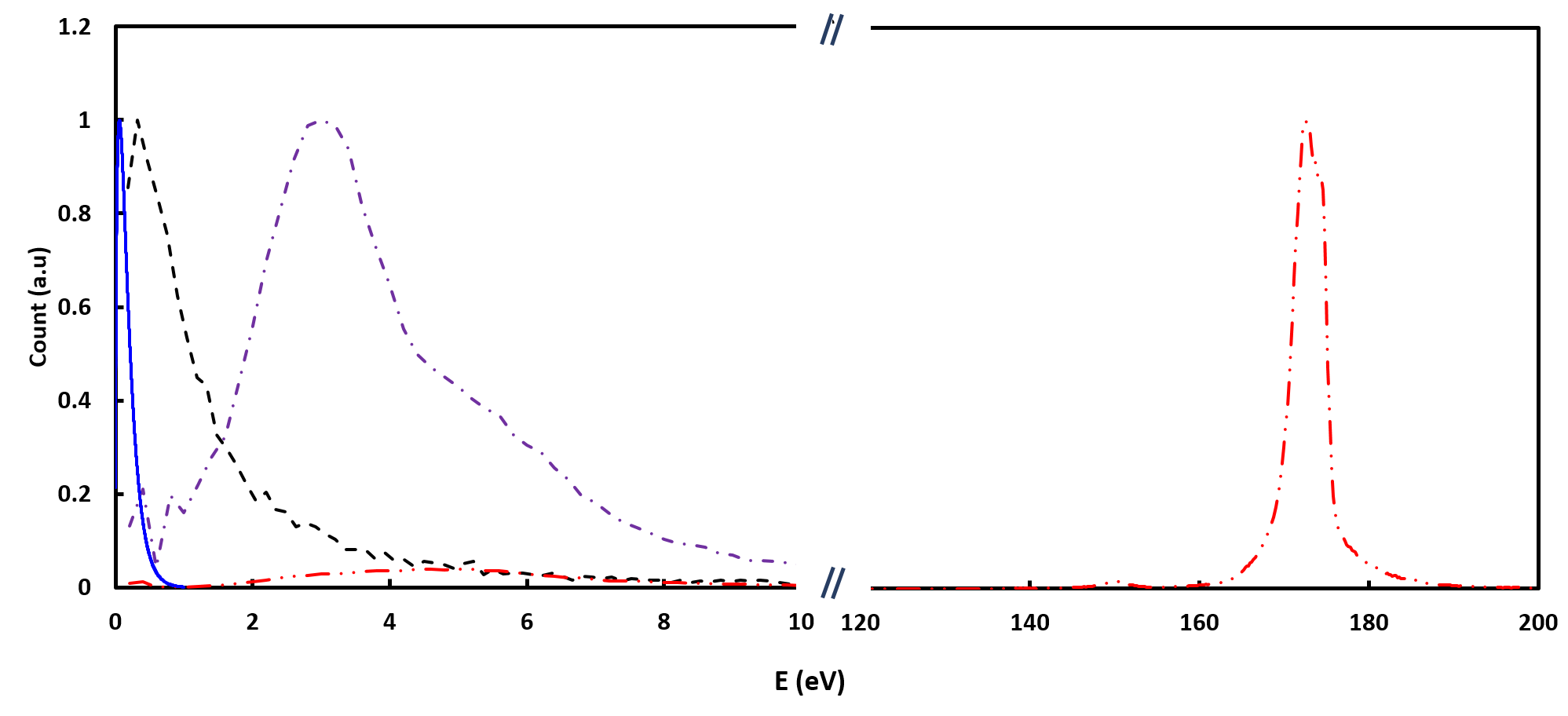}
 \caption{Sputtered Cu atom and ion energy distributions (linear scale). — Blue: Thermal Evaporation at 1358 K, - - - black: dcMS SRIM propagated along 10~cm target to substrate path, at 0.7~Pa, ‒ $\bullet$ ‒ purple:  HIPIMS Cu$^+$, Exp QMS at 10~cm, 0.7~Pa, 
 - $\bullet$ $\bullet$ -  red: bipolar HiPIMS (+180~V).}
 \label{fig:ieds}
\end{center}
\end{figure}
For thermal evaporation, a Maxwell-Boltzmann distribution at melting temperature is considered; for dcMS, SRIM AED modified by transport through the MS reactor is used \citep{xie14:224004}. For HiPIMS and bipolar HiPIMS, the kinetic energy of the sputtered ions was selected from the experimental energy-resolved mass spectra (Figure \ref{fig:ieds}), while that of sputtered neutral Cu was assumed similar to dcMS, and were selected in the corresponding AED. The constant kinetic beam-like Cu energy
was chosen as the mean kinetic energy of the SRIM AED propagated through 10 cm at 0.7 Pa.
 Figure \ref{fig:ieds} shows the different AEDs and IEDs used in this study.
The Embedded Atom Method was chosen to describe Cu, Fe, Cr, Ni, and Mo interactions \citep{daw84:6443,daw93:251}. Cross
interactions use the Johnson mixing rule \citep{johnson89:12554}.

\begin{figure*}[htp!]
\begin{center}
\includegraphics[width=4.6 in]{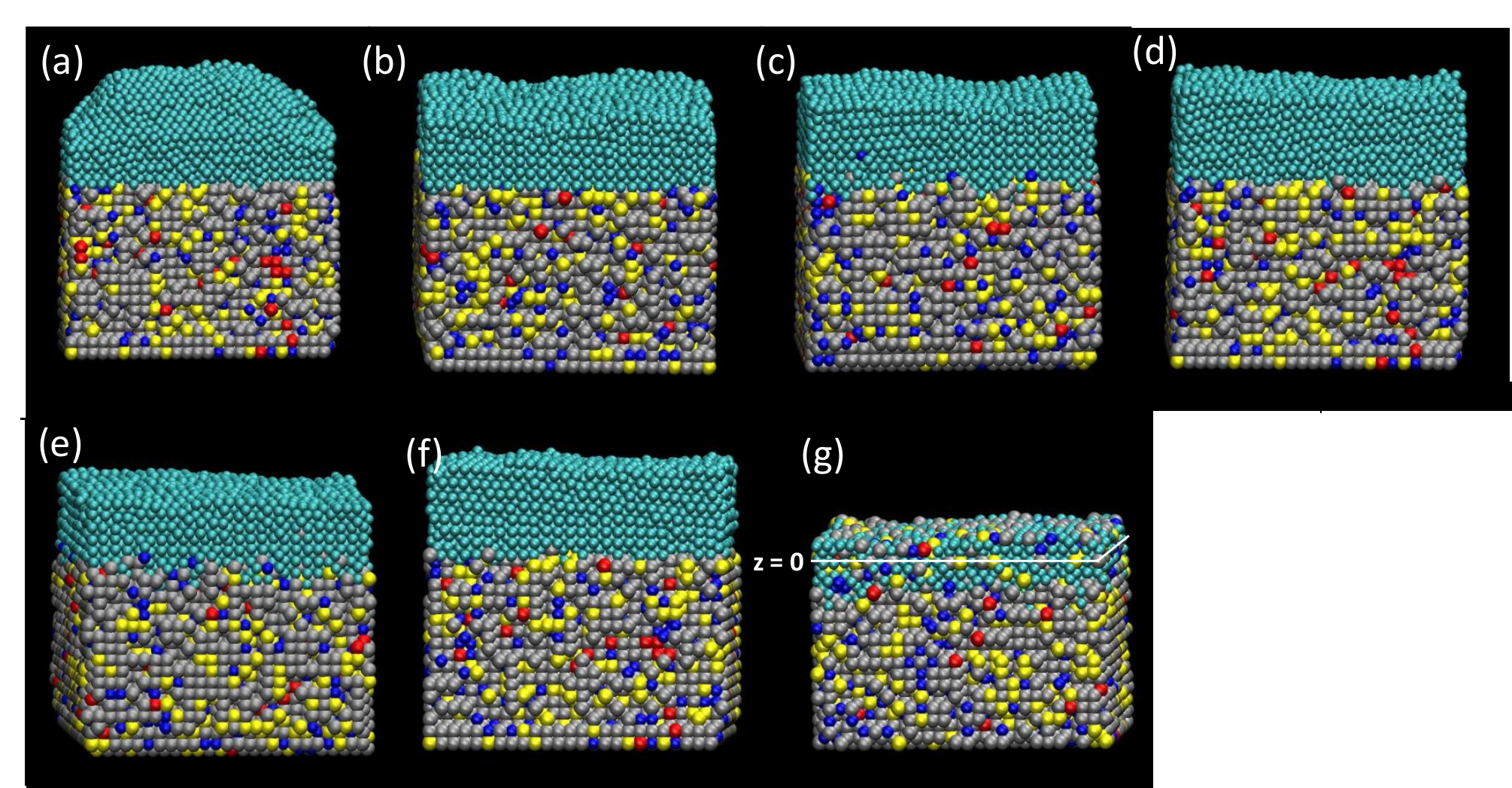}
 \caption{  Snapshots of the simulated films after 40 ns. (a) Cu beam like at ${\cal E}_{\rm n} = 1.45$ eV, (b) thermal evaporation, (c) dcMS, (d) HIPIMS 90\% Cu-10\% Cu$^+$, (e) bipolar HIPIMS +180 V 90\% Cu-10\% Cu$^+$, (f) HIPIMS 50\% Cu-50\% Cu$^+$, and (g) bipolar HIPIMS +180 V 50\% Cu-50\% Cu$^+$. The location indicated as $z=0$ is the initial substrate surface. Color
code: cyan Cu, yellow Cr, silver Fe, red Mo, blue Ni. }
 \label{fig:snapshots}
\end{center}
\end{figure*}
Figure \ref{fig:snapshots} shows the resulting deposited films with different initial conditions, mimicking the different
deposition processes. Visual inspection of the final snapshots shows that the  films deposited by constant-energy neutral beam, thermal evaporation, and dcMS have rougher surfaces (Figures \ref{fig:snapshots} (a) -- (c)), as expected \citep{kateb19:031306}.  The latest top surfaces were obtained using bipolar HiPIMS (+180 V) with   10\% Cu$^+$ (Figure \ref{fig:snapshots} (e)) and bipolar  HiPIMS (+180 V) with 50\%  Cu$^+$ (Figure \ref{fig:snapshots} (f)).
The more aggressive deposition condition, bipolar HiPIMS, results in high interface mixing and a low Cu content, suggesting Cu$^+$
sputtering of the depositing film, since the  acceleration voltage was 180 V. This was not visible in the 10\% Cu$^+$
conditions owing to the low Cu$^+$ flux.

Table \ref{table1} summarizes the properties of the simulated films. The atomic structures of the films were calculated using the Polyhedral Template Matching method \citep{larsen16:055007} available in the OVITO software \citep{stukowski10:015012}.
\begin{table*}
\caption{The Cu film properties derived from the simulated films.   \label{table1}}
\begin{tabular}{lcccc} \hline \hline 
&   Cu Sticking & Substrate erosion & Interface & Film structure  \\
& coefficient & rate (\%) & height (nm) & (\% fcc)\\
\hline 
Monoenergetic Cu atom beam & 0.98 &  $<$ 0.01 & 0 & 60 \\
Thermal evaporation & 0.76 & 0 &0 &50 \\
dcMS & 0.86 &  0.3 & 0.5 &  48   \\
HiPIMS 10\% Cu$^+$ & 0.91 &0.01& 0.3& 54 \\
Bipolar +180 V HiPIMS 10\% Cu$^+$ & 0.79 &0.5& 1.5 &57 \\
HiPIMS 50\% Cu$^+$ & 0.94 &0 &0.5& 42 \\
Bipolar +180 V HiPIMS 50\% Cu$^+$ & 0.42& 1& 5& 8 \\   \hline \hline
\end{tabular}
\end{table*}
The highest structural film quality (60\% fcc) was achieved by the monoenergetic neutral beam with a kinetic energy corresponding to the dcMS mean kinetic energy (1.45 eV) 10 cm away from the target at 0.7 Pa. However, this condition is challenging to reproduce experimentally. The next highest crystallinity was obtained for bipolar HiPIMS with 10\% Cu$^+$, also providing a non-negligible mixing interface that is expected to improve film adhesion (Figure \ref{fig:snapshots} (e)). Regular HiPIMS without a secondary positive pulse showed a similar film quality, but the interface mixing was reduced (Figure \ref{fig:snapshots} (d)). Both had high sticking coefficients. Thermally evaporated and dcMS-deposited films exhibited similar film quality (50\% fcc), but thermal evaporation showed a sticking coefficient that was 13\% lower.
The high ion rate in bipolar HiPIMS allows for large layer mixing, but at the expense of low Cu sticking owing to
film self-sputtering \citep{hong07:70}. Additionally, it exhibited the largest substrate erosion rate (1\%), whereas it was
marginal for all the other processes.

Atomic stress is an important property for comparing films deposited by the different processes \citep{hong10:520,windischmann92:547}. The calculated atomic virial stress \citep{shen04:91}  is expressed as:
\begin{equation} 
\sigma_i = - m_i \frac{{\rm d}}{{\rm d} t} \vec{u}_i \times \frac{{\rm d}}{{\rm d} t} \vec{u}_i + \frac{1}{2} \sum_{j \neq i} \vec{r}_{ij} \times \vec{f}_{ij} 
\end{equation}
 where, $i$ and $j$ are atom indices and  $\vec{f}_{ij}$ is the interatomic force between atoms $j$ and $i$. The summation is over all neighboring atoms within the force cutoff, $m_i$ is the mass of atom $i$, $\vec{u}_i$ is the displacement vector of atom $i$ relative to a reference 
 position, $\vec{r}_{ij} = \vec{r}_{j} - 
 \vec{r}_{i}$, and $\times$ is the tensor 
 product of the two vectors.

\begin{figure}[htp!]
\begin{center}
\includegraphics[width=3.1 in]{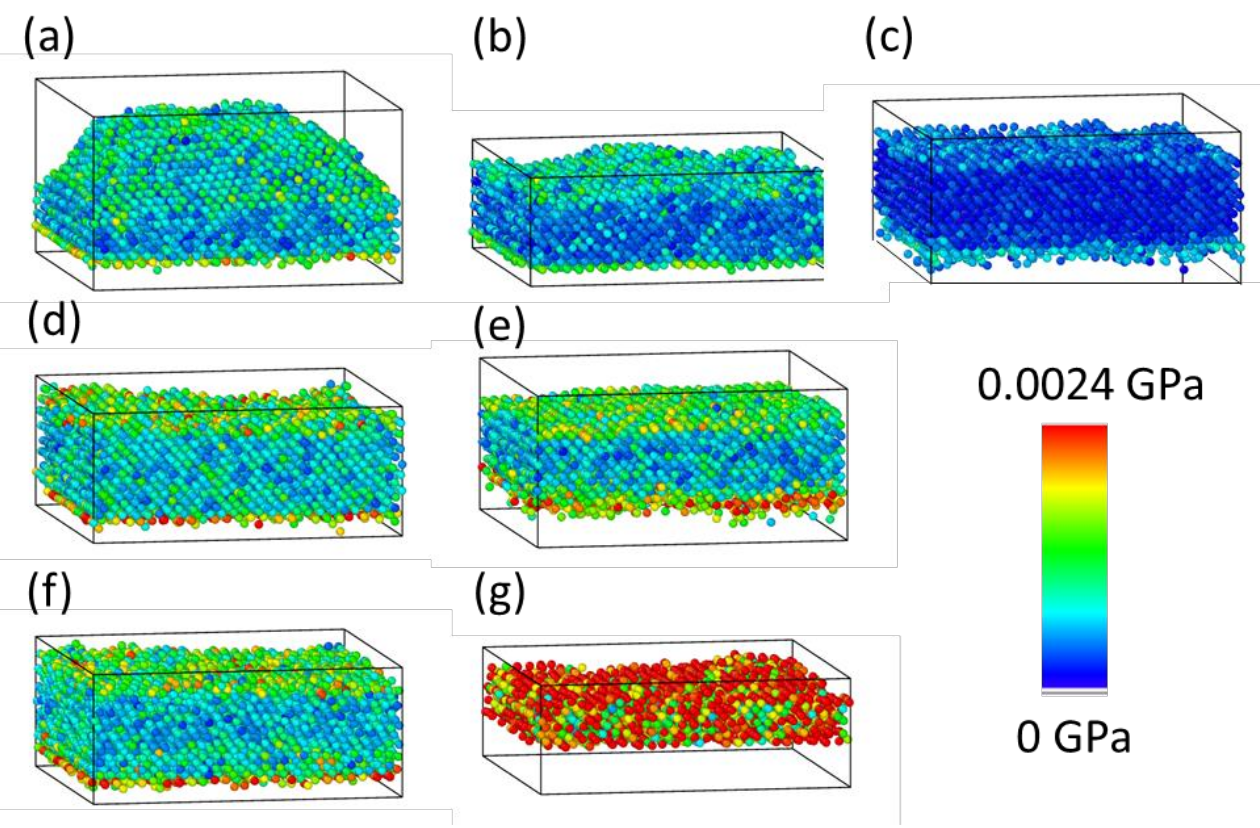}
 \caption{ Atomic virial stress in  the films deposited via the various PVD processes. (a) Cu beam like at ${\cal E}_{\rm n} = 1.45$ eV, (b) thermal evaporation, (c) dcMS, (d) HIPIMS 90\% Cu-10\% Cu$^+$, (e) bipolar HIPIMS +180 V 90\% Cu-10\% Cu$^+$, (f) HIPIMS 50\% Cu-50\% Cu$^+$, and (g) bipolar HIPIMS +180 V 50\% Cu-50\% Cu$^+$. }
 \label{fig:stress}
\end{center}
\end{figure}
In Figure \ref{fig:stress} the stress magnitude of each atom (represented by a sphere) is demonstrated by the colour map for
the films drawn in Figure \ref{fig:snapshots}. The substrate has been removed for clarity. The dcMS deposited film (Figure \ref{fig:stress} (c)) exhibits the lowest stress while the highest stress is obtained for the +180 V bipolar HiPIMS deposited film with 50\% Cu$^+$
(Figure \ref{fig:stress} (g)).

 The beam-like deposition (Figure \ref{fig:stress} (a)) did not display a high stress, but the roughness was the highest. The thermal evaporation film exhibits low stress (Figure \ref{fig:stress} (b)). 
 The three HiPIMS deposited films shown in Figure \ref{fig:stress} (d), (e) and (f), provide reasonably low stress compared to the dcMS deposited film. It should be noted that more
stressed atoms are present at the interface with the substrate. The bipolar +180 V 50\% Cu$^+$ film (Figure \ref{fig:stress} (e)) 
shows less stress in the inner film than the other HiPIMS deposited films. This means that bipolar HiPIMS can provide a relaxed film. This has to be linked to the highest degree of crystal ordering, as shown in Table 1, where the fcc fraction reached 57\%.
 
 In summary, molecular dynamics simulations of various PVD processes can be handled and differentiated using the corresponding AEDs and IEDs of sputtered atoms and ions. It is possible to use the experimental IEDs obtained by energy-resolved mass spectrometry. This allows for a better account of the process characteristics. In the present case of Cu deposition, bipolar +180 V HiPIMS with 10\% Cu$^+$ ions
exhibited the best film properties in terms of crystallinity and atomic stress among the PVD processes investigated.


\begin{acknowledgments} 


{\bf Data Availability Statement}
Research data supporting this publication are available from the author on
reasonable request.

{\bf Conflict of interest} 
 The authors declare that they have no affiliations with or involvement in
any organization or entity with any financial interest in the subject matter or materials discussed in this
manuscript.

\end{acknowledgments}


\bibliography{heim78}
\bibliographystyle{apsrev4-1}

%
%

%
%

\end{document}